\documentclass[epj,nopacs]{svjour}
\usepackage{graphicx}
\newcommand{\GeV}{\textrm{GeV}}

\begin{document}
\title{Measurements of Event and Jet Shapes at HERA}
\author{Thomas Kluge
\thanks{On behalf of the H1 and ZEUS Collaborations}%
}                     
%
%
\institute{I. Physikalisches Institut, RWTH Aachen, D-52056 Aachen, Germany}
\date{Received: October 31, 2003}
%
\abstract{
Event shapes and jet shapes in neutral current deep inelastic scattering and photoproduction are studied with the H1 and ZEUS detectors at HERA.
The measurements are compared to next-to-leading-order QCD calculations and Monte Carlo models.
The strong coupling constant is determined from subjet multiplicities.
} 
\maketitle

\section{Introduction}
The HERA collider is a precision tool for the study of QCD.
The H1 and ZEUS Collaborations have performed a wealth of analyses which investigate the ha\-dro\-nic final state in $ep$ collisions.
In addition to the measurement of plain jet cross sections, it is rewarding to study the substructure of jets, or in the case of event shapes, the structure of the whole hadronic final state.
In this case, uncertainties connected to the determination of cross sections for jet production do not compromise the results.\\
In the regime of high transverse momenta, experimental uncertainties and non-perturbative effects are small, which makes possible precision tests of perturbative QCD and the extraction of the strong coupling constant.\\
On the other hand, understanding the nature of hadronisation is a challenge, which is important for all high energy experiments involving hadrons.
In this area, the concept of power corrections offers prospects beyond phenomenological models implemented in Monte Carlo generators.  
\label{intro}

\section{Subjet Multiplicities}
\label{subjets}
\begin{center}
\begin{figure}
\resizebox{0.4\textwidth}{!}{%
\includegraphics{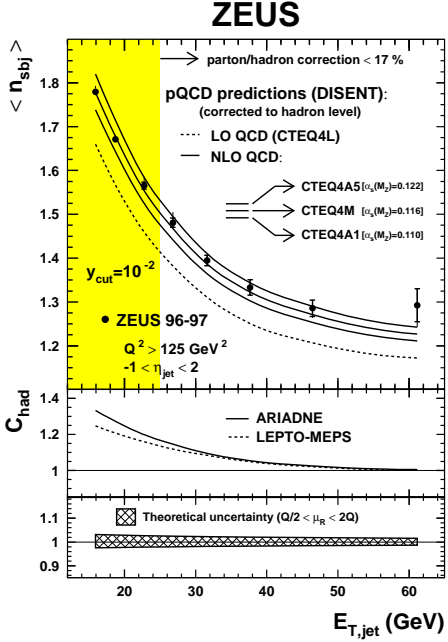}
}
\caption{The mean subjet multiplicity of an inclusive jet sample in NC DIS, corrected to hadron level, as a function of $E_{T\mathrm{jet}}$.}
\label{subjetsfig}
\end{figure}
\end{center}

The internal structure of jets is determined by QCD radiation processes.
Non-perturbative fragmentation effects become small at high transverse jet energies $E_T$ and values of the resolution scale $y_\mathrm{cut}$ that are not too low.
Perturbative QCD calculations can be applied. In this region of phasespace they depend only weakly on parton density functions, but are directly proportional to the strong coupling constant $\alpha_s$.\\
The mean subjet multiplicity is defined as 
\begin{displaymath}
\langle n_\mathrm{sbj}(y_\mathrm{cut})\rangle=\frac{1}{N_\mathrm{jets}}\sum_{i=1}^{N_\mathrm{jets}}n^i_\mathrm{sbj}(y_\mathrm{cut}),
\end{displaymath}
where $n^i_\mathrm{sbj}(y_\mathrm{cut})$ is the number of subjets in jet $i$ and $N_\mathrm{jets}$ is the total number of jets in the sample.\\
The ZEUS collaboration has measured the mean subjet multiplicity in neutral current deep inelastic scattering (NC DIS) as a function of $E_T^{\mathrm{jet}}$\cite{Chekanov:2002ux}.
Jets were defined by using the longitudinally invariant $k_T$ cluster algorithm in the laboratory frame.
In Fig.~\ref{subjetsfig} the measurements of $\langle n_\mathrm{sbj} \rangle$ are compared with the NLO prediction from DISENT\cite{Catani:1997vz}, corrected for hadronisation efects.
The overall description of the data is good, so that the measurements can be used to make a determination of $\alpha_s$.
The region used for the fit was restricted to $E_T^{\mathrm{jet}}>25\GeV$, where hadronisation corrections differ by less than $17\%$ from unity.\\
The QCD fit of the mean subjet multiplicity for $25<E_T^{\mathrm{jet}}<71~\GeV$ at $y_\mathrm{cut}=10^{-2}$ yields
\begin{displaymath}
\alpha_s(m_Z)=0.1187\pm0.0017(\textrm{stat.})^{+0.0024}_{-0.0009}(\textrm{syst.})^{+0.0093}_{-0.0076}(\textrm{th.}).
\end{displaymath}
This result is consistent with recent determinations by the H1 \cite{Adloff:2000tq,Adloff:2000qk} and ZEUS \cite{Chekanov:2002be,Chekanov:2002pv} Collaborations and with the PDG value \cite{Hagiwara:2002fs}.
The total error is dominated by the theoretical uncertainty due to missing terms beyond NLO. 

\section{Substructure Dependence of Jet Cross Sections}
\label{jetsubstructure}
\begin{center}
\begin{figure}
\resizebox{0.5\textwidth}{!}{
\includegraphics{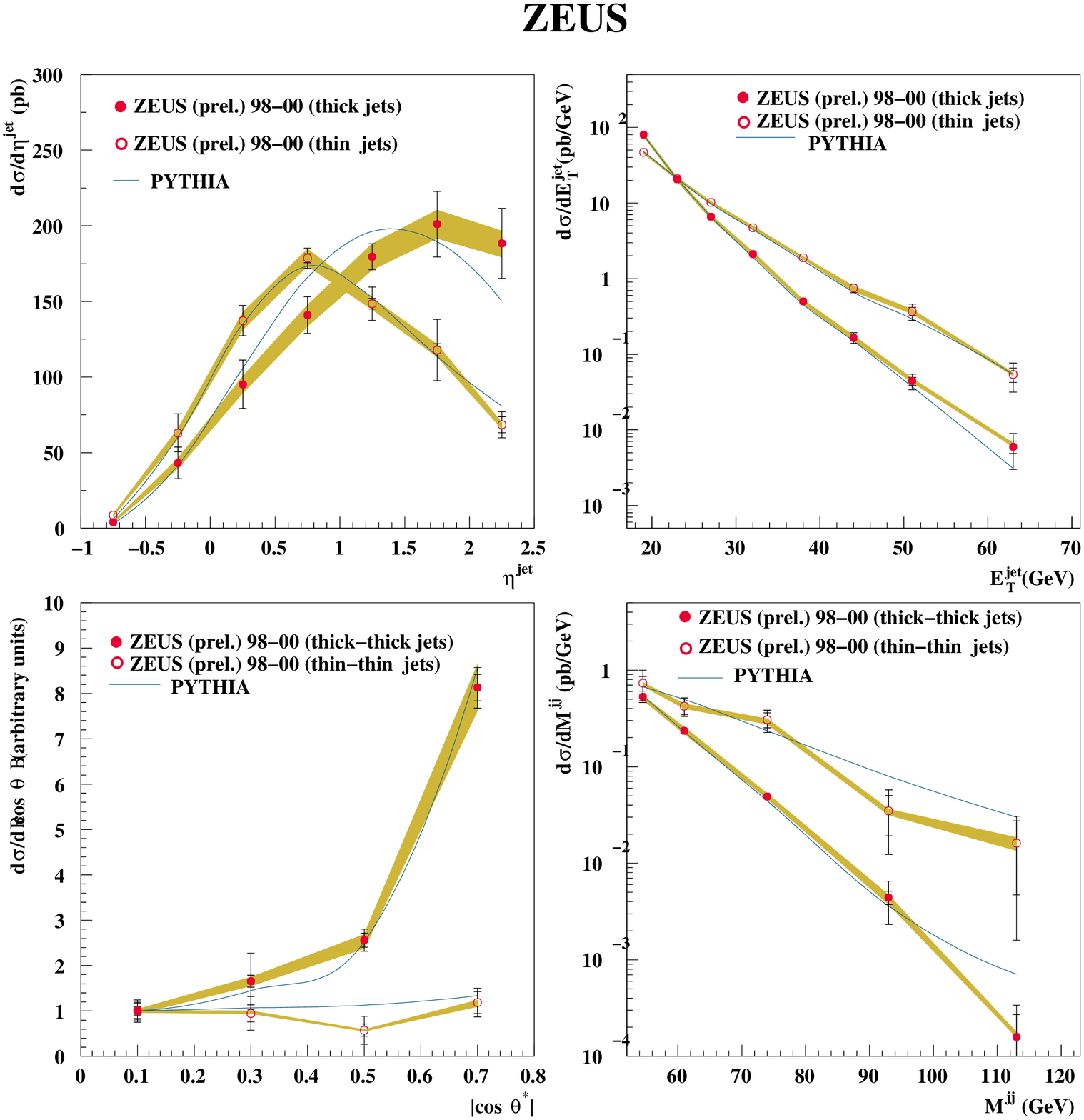}
}
\caption{Differential jet cross section in photoproduction for an inclusive (upper) and dijet (lower) sample.
 The jets were assigned to ``thick'' and ``thin'' subsamples according to their shape and subjet multiplicity.}
\label{jetsubstructurefig}
\end{figure}
\end{center}

Quarks and gluons carry different colour charges, therefore the internal structure of a jet is expected to depend on the type of the primary parton from which it originated.
Obervables which probe the internal structure are the subjet multiplicity (section \ref{subjets}) and the integrated jet shape $\Psi(r)$,  which is defined as 
\begin{displaymath}
\Psi(r)=\frac{E_T(r)}{E_T^\mathrm{jet}},
\end{displaymath}
where $E_T(r)$ is the transverse energy within a cone of radius $r=\sqrt{\Delta\phi^2+\Delta\eta^2}$ about the jet axis.\\
In photon-proton reactions, two types of processes contribute to jet production.
In the direct process the photon interacts with a parton in the proton as a whole, whereas in the resolved process it behaves as a source of partons which subsequently interact with partons from the proton.
Jets from direct and resolved processes exhibit different angular distributions and $E_T$ spectra.
Consequently, dijets from both processes differ in their distributions of invariant mass $M^{jj}$ and the angle between the jet-jet axis and the beam direction in the dijet centre-of-mass system $\cos \theta^{\ast}$. 
The cross section for subprocesses with gluons in the final state is larger for the resolved than for the direct process, if an appropriate region of phase space is chosen.
Therefore, the experimental investigation of jet substructure as a function of jet kinematics allows one to study the characteristics of quark- and gluon-initiated jets.\\
The ZEUS Collaboration has measured differential inclusive jet and dijet cross sections in photoproduction \cite{ZEUS:2003eps518}.
The internal structure of jets, in terms of the integrated jet shape and subjet multiplicity, has been used to select samples enriched in quark- and  gluon-initiated jets, classified as ``thin'' 
and ``thick'' samples, respectively.
Fig.~\ref{jetsubstructurefig} shows the measured inclusive jet cross section as a function of $E_T$ and $\eta_\mathrm{jet}$, and the dijet cross section as a function of  $|\cos \theta^{\ast}|$ and $M^{jj}$.
The distributions are in general well described by calculations using PYTHIA, which have been  normalised to the total  measured cross section.
The ``thick'' data sample behaves according to expectations for gluon initiated jets in the ivestigated phase space: 
these jets are more forward, have a softer $E_t$ spectrum (inclusive jets), lower invariant mass and the jet axis is closer to the beam direction (dijets). 
This behaviour is consistent with the expected dominance of the resolved subprocess for gluon initiated jets. 

\section{Event Shapes}
\label{eventshapes}
\begin{figure}
\resizebox{0.5\textwidth}{!}{%
\includegraphics{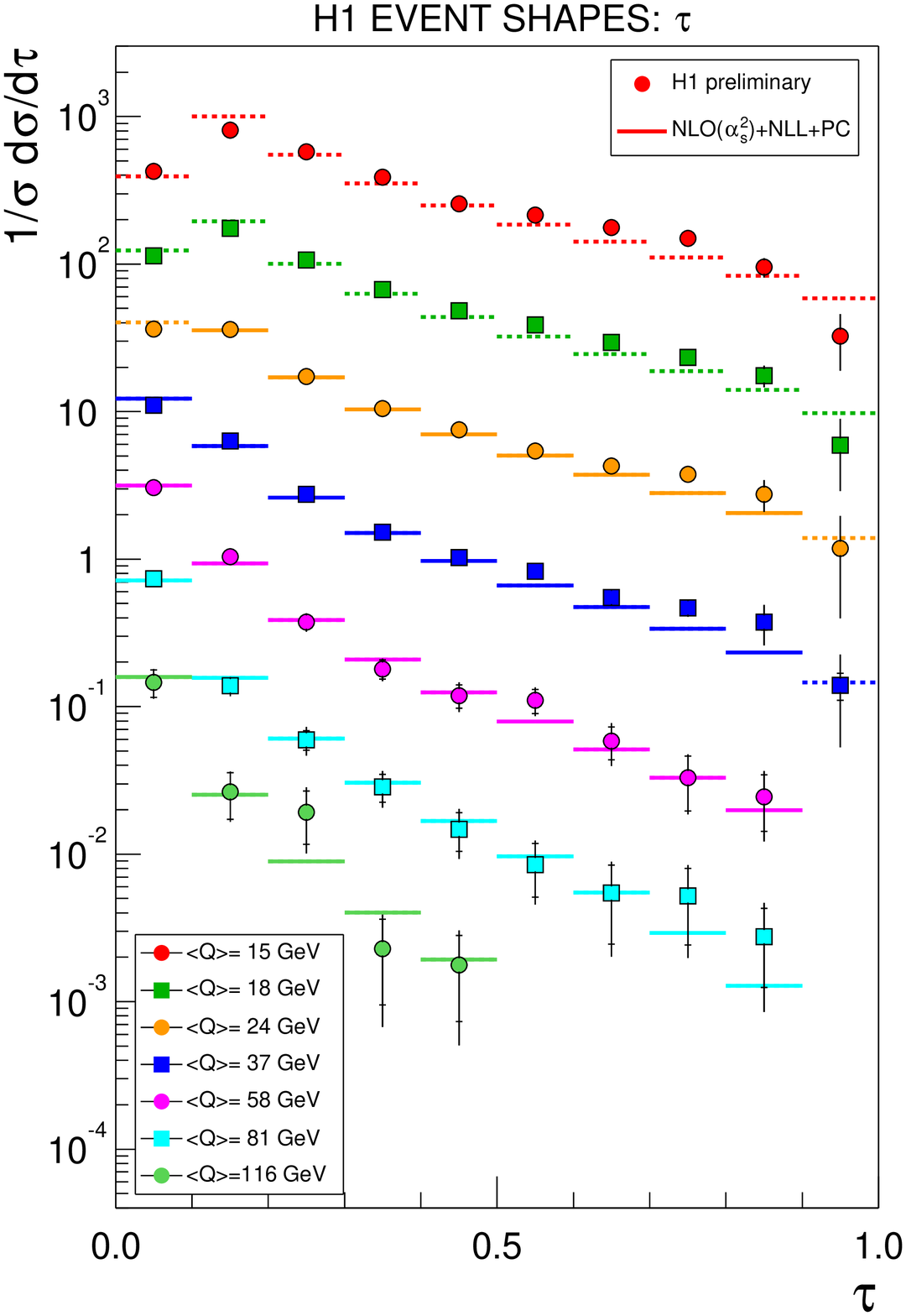}
\includegraphics{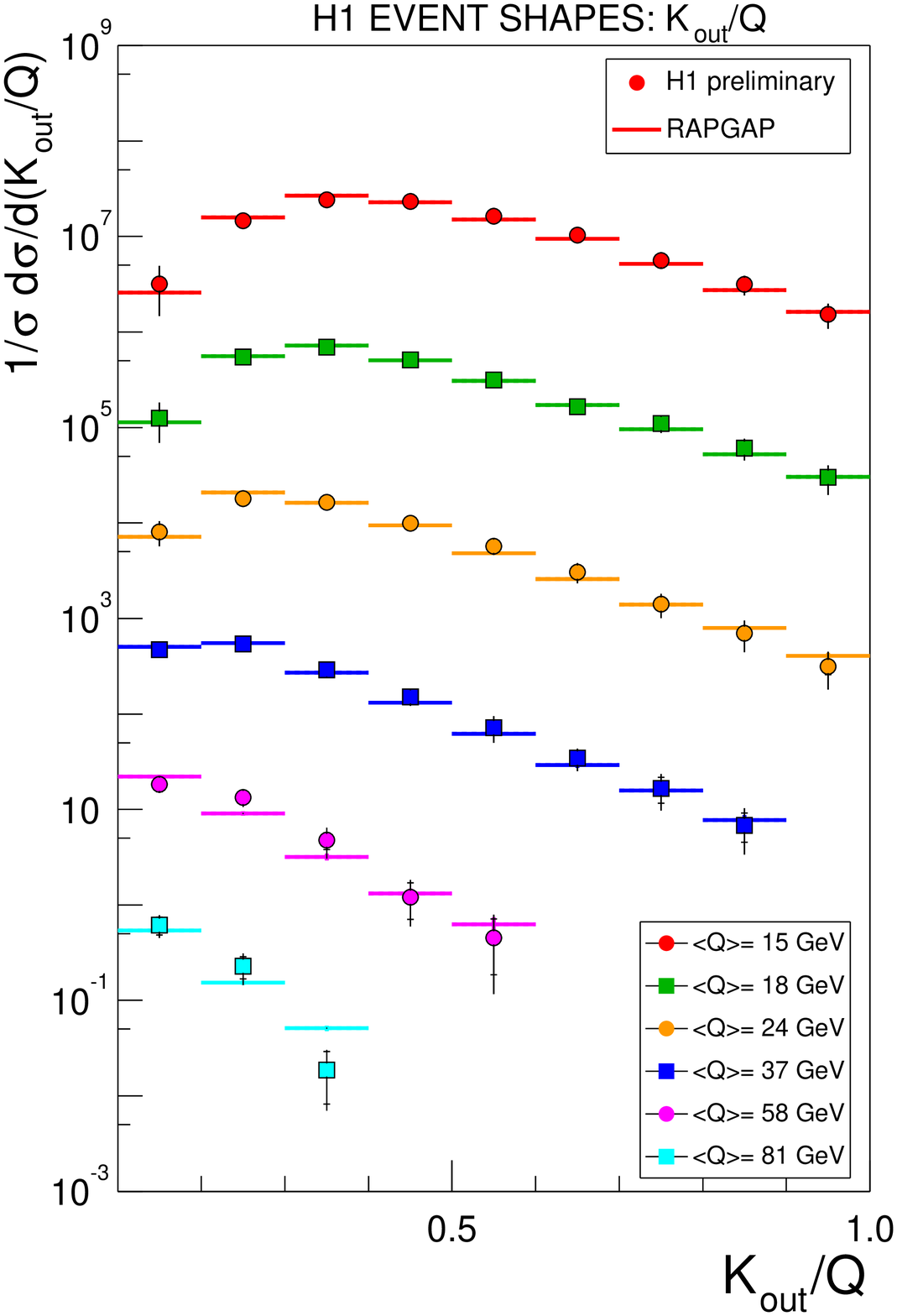}
}
\caption{Measured distributions of 1--thrust $\tau$ (left) and out-of-plane momentum $K_\mathrm{out}$ (right).
  $\tau$ is compared with the results of a fit based on NLO QCD including resummation and power corrections,
 $K_\mathrm{out}$ is compared with results from the RAPGAP Monte Carlo generator.}
\label{eventshapedist}
\end{figure}

\begin{figure}
\resizebox{0.5\textwidth}{!}{%
\includegraphics[height=4.8cm]{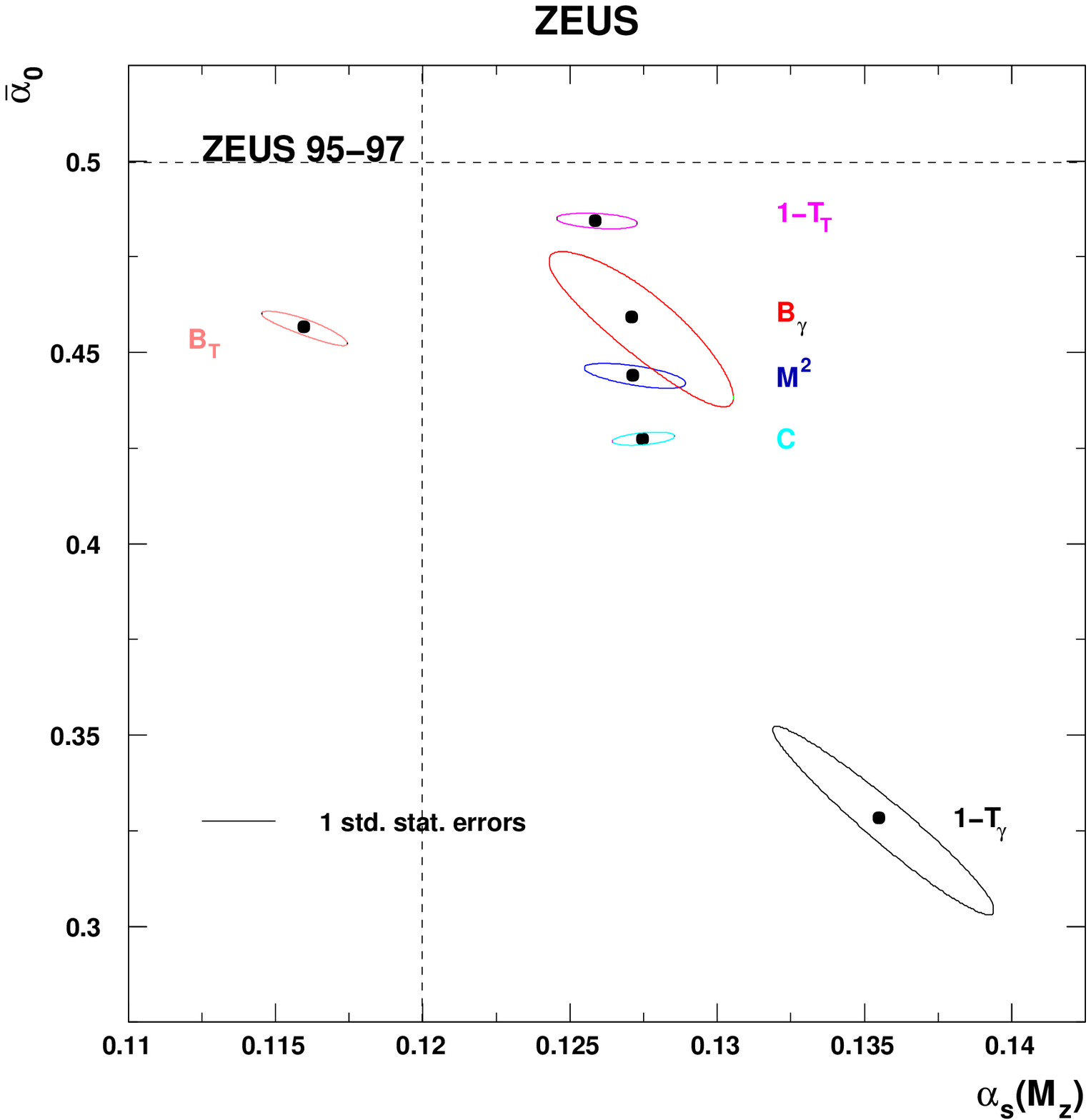}
\includegraphics[height=5cm]{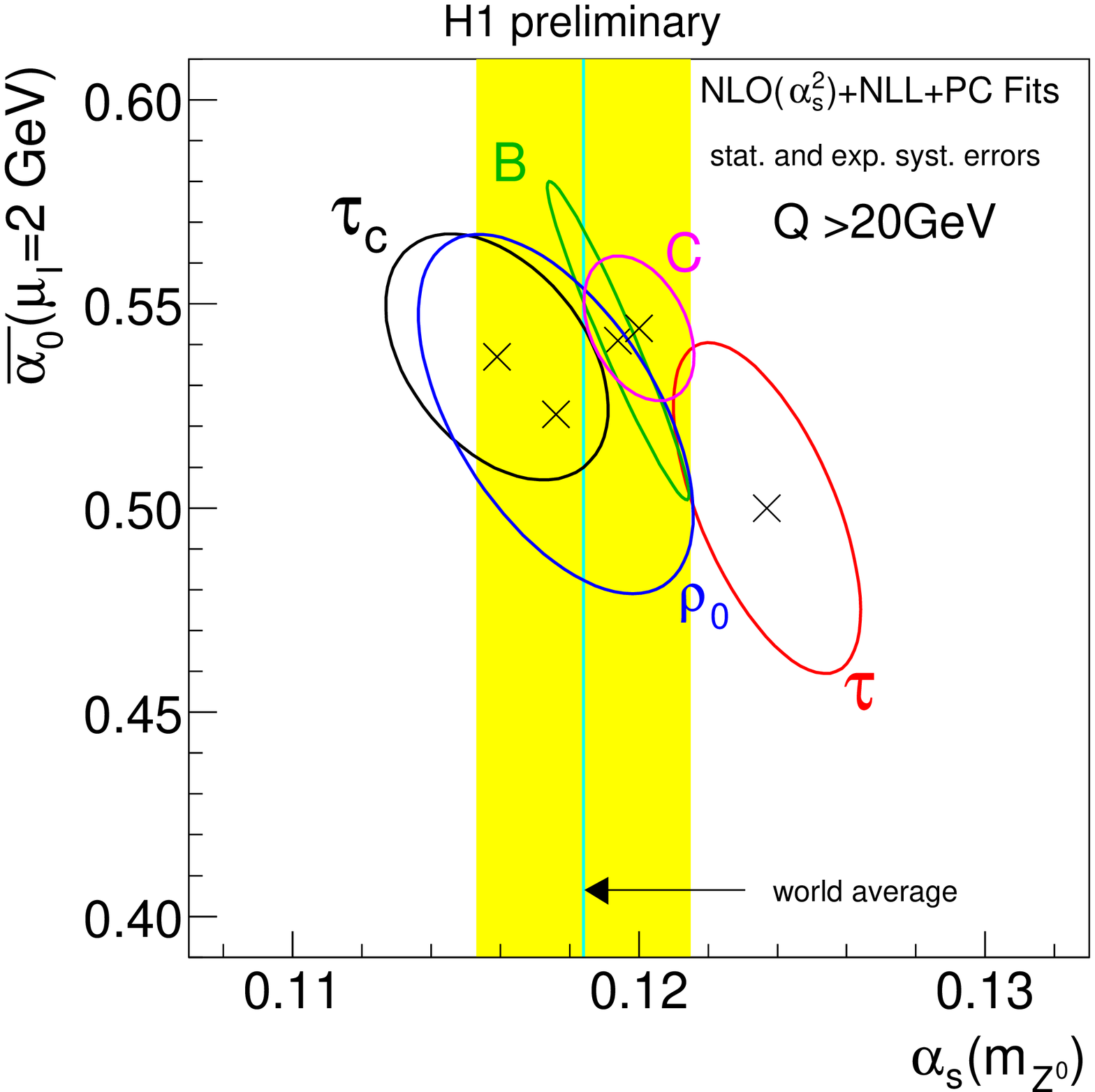}
}
\label{eventshapefits}
\caption{1--$\sigma$ contours in the $(\alpha_s,\bar \alpha_0)$ plane from fits to event shape mean values (left) and differential distributions (right). }
\end{figure}

Event shape variables are designed to study QCD by measuring properties of the flow of hadronic energy-mo\-men\-tum.
A suitable frame of reference in DIS is the Breit frame, which divides an event into hemispheres corresponding to the proton remnant and the hadronic final state evolving from the struck parton.
Event shape observables are then usually defined for the particles in the current hemisphere (CH) alone, thereby avoiding non-perturbative effects connected with the proton remnant.
A commonly used event shape variable is thrust $\tau\equiv 1-T$:
\begin{displaymath}
T=\frac{\sum_{h\epsilon\mathrm{CH}}|p_{zh}|}{\sum_{h\epsilon\mathrm{CH}}|p_h|},
\end{displaymath}
where $p_h$ is the momentum of particle $h$ in the current hemisphere and $p_{zh}$ this momentum projected on the axis of the exchanged virtual photon.  
The description of such infrared-collinear safe observables by fixed order calculations faces two difficulties.
Firstly, non-perturbative hadronisation effects can be large, even at formally perturbative scales.
In the approach by Dokshitzer, Webber et al. \cite{Dokshitzer:1997ew} these effects can be addressed using power corrections proportional to $1/Q$, governed by one universal free parameter $\alpha_0$.
Secondly, the convergence of the perturbative series at low values of the event shape variables is very poor, making a resummation to all orders in $\alpha_s$ necessary.
Resummed NLL calculations, matched to NLO, have recently become available for two kinds of thrust, $\tau_c$ and $\tau$, the jet broadening $B$, the jet mass $\rho_0$ and the C-parameter\cite{Dasgupta:2002dc}.
While these event shapes are mainly sensitive to the transition from 1-jet like to 2-jet like topologies, there are now also measurements and calculations available for the out-of-plane momentum $K_{\mathrm{out}}$ and the azimuthal correlation $\chi$, which probe the transition from  2-jet to 3-jet topologies \cite{Banfi:2001ci}. \\
Mean values of event shapes as a function of $Q$ have been studied by the H1 \cite{Adloff:1999gn} and ZEUS \cite{Chekanov:2002xk} Collaborations.
The H1 Collaboration has also measured spectra of event shape variables \cite{H1:2003eps111}.
Fig.~\ref{eventshapedist} shows the distributions of thrust and out-of-plane momentum in bins of momentum transfer $Q$.
The thrust spectra are compared with the results of a fit based on resummed calculations matched to NLO and including power corrections.
For the out-of-plane momentum the matching is not yet available, therefore a comparison with the RAPGAP Monte Carlo event generator \cite{Jung:1995gf} is presented.
This includes LO matrix elements and parton showers.  
Both observables are well described over the whole region of phase space.\\
The ZEUS Collaboration has made a fit to the mean values of event shape variables using DISASTER++ \cite{Graudenz:1997gv} and the Dok\-shitzer-Web\-ber power corrections, as shown in Fig.~\ref{eventshapefits} (left). The contours correspond to 1-$\sigma$ statistical errors.
Consistent values of $\alpha_s$ are obtained for four of the shape variables with $\alpha_0$ values that agree within $10\%$.
However, the lack of overall consistency in  $\alpha_s$ and $\alpha_0$ suggests that higher order processes which are not included may be important.
On the right hand side of Fig.~\ref{eventshapefits} results of a fit to event shape spectra by the H1 Collaboration are shown, including resummed calculations.
Here the contours indicate 1-$\sigma$ statistical and experimental systematic errors, added in quadrature.
The results for the five studied event shapes are consistent with each other and with the world average of $\alpha_s$ \cite{Bethke:2000ai}. The universal non-perturbative parameter $\alpha_0$ is confirmed to be 0.5 at the $10\%$ level. 

\section{Conclusion}
\label{conclusion}
Differential distributions of jet and event shapes have been measured by the H1 and ZEUS Collaborations with high precision. 
Calculations based on perturbative and non-perturbative QCD describe the measurements over a large region of phasespace.
 
%
\bibliographystyle{h-physrev}
\bibliography{tkbib}

\end{document}